\documentclass[12pt]{article}

\usepackage{amsmath,amssymb}
\usepackage{bm}
\usepackage{graphicx, color}
\usepackage{wrapfig}
\usepackage{cite}
\usepackage[utf8]{inputenc}
\usepackage{a4wide}
\usepackage{amsfonts}
\usepackage{color}
\usepackage{adjustbox}
\usepackage{multirow}
\usepackage{pdflscape}
\usepackage{textcomp}
\usepackage{gensymb}
\usepackage{graphicx}
\graphicspath{{figures/}}
\usepackage{diagbox}
\usepackage{pifont}
\usepackage{bigstrut}
\usepackage[small,bf]{caption}
\usepackage{subcaption}
\usepackage{tabulary}
\usepackage{booktabs}
\usepackage{stackengine}
\usepackage{mathtools}
\usepackage{enumerate}
\usepackage[unicode]{hyperref}

\newcommand{\be}{\begin{equation}}
\newcommand{\ee}{\end{equation}}

\def\1{\mathbf{1}}

\def\2{\mathbf{2}}
\def\3{\mathbf{3}}

\def\g{\gamma}
\def\G{\Gamma}

\def\t{\tau}

\def\ltap{\ \raisebox{-.4ex}{\rlap{$\sim$}} \raisebox{.4ex}{$<$}\ }

\DeclareMathOperator{\diag}{diag}

\DeclareMathOperator{\re}{Re}

\DeclarePairedDelimiter{\vevdel}{\langle}{\rangle}
\newcommand{\vev}{\vevdel}
\allowdisplaybreaks
\numberwithin{equation}{section}
\makeatletter
\g@addto@macro\bfseries{\boldmath}
\makeatother

\begin{document}
	\title{
		\begin{flushright}
			\begin{minipage}{0.3\linewidth}
				\normalsize
				SISSA 57/2018/FISI\\
				IPMU18-0209\\*[50pt]
			\end{minipage}
		\end{flushright}
		{\Large \bf 
			Trimaximal Neutrino Mixing from
			Modular $A_4$ Invariance with Residual Symmetries
			\\*[20pt]}}
	
	\author{ 
		\centerline{
			~P. P. Novichkov $^{a}\footnote{E-mail address: pavel.novichkov@sissa.it}$~,
			~~S. T. Petcov $^{a,b}\footnote{Also at
				Institute of Nuclear Research and Nuclear Energy,
				Bulgarian Academy of Sciences, 1784 Sofia, Bulgaria.}$~,
			~~M. Tanimoto $^{c}\footnote{E-mail address: tanimoto@muse.sc.niigata-u.ac.jp}$} \\*[5pt]
		\centerline{
			\begin{minipage}{\linewidth}
				\begin{center}
					$^a${\it \normalsize
						SISSA/INFN, Via Bonomea 265, 34136 Trieste, Italy} \\*[5pt]
					$^b${\it \normalsize
						Kavli IPMU(WIP), University of Tokyo, 5-1-5 Kashiwanoha, 277-8583 Kashiwa, Japan} \\*[5pt]
					$^c${\it \normalsize
						Department of Physics, Niigata University, Niigata 950-2181, Japan}
				\end{center}
		\end{minipage}}
		\\*[50pt]}
\date{
\centerline{\small 
\bf Abstract}
\begin{minipage}{0.9\linewidth}
\medskip 
\medskip 
\small 
We construct phenomenologically viable models of lepton masses and 
mixing based on modular $A_4$ invariance broken to residual 
symmetries $\mathbb{Z}^{T}_3$ or $\mathbb{Z}^{ST}_3$ 
and $\mathbb{Z}^S_2$ respectively in the charged lepton and 
neutrino sectors. In these models the neutrino mixing matrix is 
of trimaximal mixing form.
In addition to successfully 
describing the charged lepton masses, neutrino mass-squared 
differences and the atmospheric and reactor
neutrino mixing angles $\theta_{23}$ and $\theta_{13}$,
these models predict the values of the lightest neutrino mass 
(i.e., the absolute neutrino mass scale), of the Dirac and 
Majorana CP violation (CPV) phases, as well as the 
existence of specific correlations between
i) the values of the solar neutrino mixing 
angle $\theta_{12}$ and the angle $\theta_{13}$ (which determines $\theta_{12}$),
ii) the values of the Dirac CPV phase $\delta$ 
and of the angles  $\theta_{23}$ and  $\theta_{13}$,
iii) the sum of the  neutrino masses and $\theta_{23}$,
iv) the neutrinoless double beta decay effective Majorana mass 
and $\theta_{23}$, and v) between the two Majorana phases.
 \end{minipage}
 }
\begin{titlepage}
\maketitle
\thispagestyle{empty}
\end{titlepage}


\section{Introduction}

 Understanding the origin of the  
flavour structure of quarks and leptons 
remains one of the outstanding problems 
in particle physics. The pattern of two large and one small 
neutrino (lepton) mixing angles, 
revealed by the data 
obtained in neutrino oscillation experiments 
(see, e.g., \cite{Tanabashi:2018oca}),
provides an important clue  in the investigations 
of the lepton flavour problem,
suggesting the existence of 
flavour symmetry in the lepton sector. 
The results of the recent global analyses 
of the neutrino oscillation data show also that a 
neutrino mass spectrum with 
normal ordering (NO) is favoured over the spectrum with inverted 
ordering (IO), as well as a preference for a value of the 
Dirac CP violation (CPV) phase $\delta$
close to $3\pi/2$ (see, e.g.,~\cite{Capozzi:2018ubv}).

  The observed 3-neutrino mixing pattern 
can naturally be explained by extending 
the Standard Theory (ST) with a flavour symmetry 
associated with a non-Abelian discrete symmetry group.
Models based on $S_3$, $A_4$, $S_4$, $A_5$ and other groups 
of larger orders have been proposed and extensively studied (see, e.g.,  
\cite{Altarelli:2010gt,Ishimori:2010au,Ishimori:2012zz,King:2013eh,King:2014nza,Tanimoto:2015nfa,Petcov:2017ggy}).
In particular, the $A_4$ flavour model attracted considerable interest 
because the $A_4$ group is the minimal one including a triplet 
unitary irreducible representation, 
which allows for a natural explanation of the  
existence of  three families of leptons 
\cite{Ma:2001dn,Babu:2002dz,Altarelli:2005yp,Altarelli:2005yx,Shimizu:2011xg,Kang:2018txu}.
In all models based on non-Abelian discrete flavour symmetry,
the flavour symmetry must be broken in order to 
reproduce the measured values of the 
neutrino mixing angles.
This is achieved by introducing typically a 
large number of ST gauge singlet scalars - 
the so-called ``flavons'' - in the 
Lagrangian of the theory,  
which have to develop a set of particularly 
aligned  vacuum expectation values (VEVs). 
Arranging for such an alignment requires the 
construction of rather elaborate scalar potentials.

  An attractive approach to the lepton flavour problem,  
based on the invariance under the modular group,
has been proposed in Ref.\cite{Feruglio:2017spp}, 
where also  models of the finite
modular group  $\Gamma_3 \simeq A_4$ have been presented.
Although the models constructed in
Ref.~\cite{Feruglio:2017spp} are not realistic and make use 
of a minimal set of flavon fields, this work 
inspired further studies of the modular invariance 
approach to the lepton flavour problem. 
The modular group includes $S_3$, $A_4$, $S_4$, and $A_5$
as its principal congruence subgroups, $\Gamma_2 \simeq S_3$,  
$\Gamma_3 \simeq A_4 $, $\Gamma_4 \simeq S_4$ 
and  $\Gamma_5 \simeq A_5$ \cite{deAdelhartToorop:2011re}. 
However, there is a significant difference between the 
models based on the modular $S_3$, $A_4$, $S_4$ etc. 
symmetry and those based on the usual 
non-Abelian discrete $S_3$, $A_4$, $S_4$ etc. flavour symmetry.
The constants of a theory based on the 
finite modular symmetry, such as Yukawa couplings and,
e.g., the right-handed neutrino mass matrix in type I seesaw 
scenario,  also transform non-trivially under the modular symmetry 
and are written in terms of modular forms which are  
holomorphic functions of a complex scalar field - 
the modulus  $\tau$. At the same time
the modular forms transform under the usual 
non-Abelian discrete flavour symmetries.
In the most economical versions 
of the models with modular symmetry, 
the VEV of the modulus $\tau$ is 
the only source of symmetry breaking 
without the need of flavon fields.

In Ref.~\cite{Kobayashi:2018vbk} a realistic model
with modular $\G_2 \simeq S_3$ symmetry 
was built with the help of a minimal set 
of flavon fields. 
A realistic model of the charged lepton 
and neutrino masses and of neutrino mixing without 
flavons, in which the modular $\G_4 \simeq S_4$ 
symmetry was used, was constructed in~\cite{Penedo:2018nmg}.
Subsequently, lepton flavour models without flavons, based on 
the modular symmetry $\G_3 \simeq A_4$ 
was proposed in Refs.~%
\cite{Criado:2018thu,Kobayashi:2018scp}. 
A comprehensive investigation of the simplest 
viable models of lepton masses and mixing, based on the modular 
$S_4$ symmetry, was performed in Ref. \cite{Novichkov:2018ovf}. 
Necessary ingredients for constructing flavour models 
based, in particular, on the modular  
symmetries $\Delta (96)$ and  $\Delta (384)$ 
have been obtained in \cite{Kobayashi:2018bff},
while for 
models based on $A_5$ symmetry they have been derived 
in \cite{Novichkov:2018xyzw}.

 If one of the subgroups of the considered finite modular group
is preserved, this residual symmetry fixes  $\tau$  to a specific 
value (see, e.g., \cite{Novichkov:2018ovf}). 
Phenomenologically viable models based on the modular $S_4$ 
and $A_5$ symmetries, broken respectively to residual $Z_3$ and 
$Z_5$ symmetries in the charged lepton sector and to a
$Z_2$ symmetry in the neutrino sector, were presented 
in Refs. \cite{Novichkov:2018ovf,Novichkov:2018xyzw}.
So far, apart from these two studies,  
the implications of residual symmetries have been 
investigated only in the framework of the  
usual non-Abelian  discrete symmetry approach to the 
lepton (and quark) flavour problem. 
It has been shown that they lead, in particular, 
to specific experimentally testable correlations 
between the values of some of the neutrino mixing angles 
and/or between the values of the neutrino mixing angles and of 
the Dirac CP violation phase in the neutrino mixing 
 \cite{Ding:2013hpa,Ding:2013bpa,Li:2014eia,Petcov:2014laa,Girardi:2015rwa,Girardi:2016zwz,Petcov:2018snn}.

 In the present article 
we construct phenomenologically viable models of lepton masses and mixing 
based on residual symmetries resulting from the breaking 
of the $A_4$  modular symmetry. 
It is found that the weight $4$ modular forms are required to 
obtain charged lepton and neutrino mass matrices leading 
to lepton masses and mixing which are consistent with 
the experimental data on neutrino oscillations.
We also find that in these models the 
PMNS matrix \cite{BPont57,Maki:1962mu,Pontecorvo:1967fh} is predicted
to be of the trimaximal mixing form 
\cite{Grimus:2008tt,Albright:2010ap}.

The paper is organized as follows.
In section 2,  we give a brief review on the modular symmetry. 
In section 3,  we discuss the residual symmetries of $A_4$ and 
their modular forms.
In section 4, we present the lepton mass matrices in the residual symmetry.
In section 5, we present models and their numerical results.
Section 6  is devoted to a summary.
Appendix A shows the relevant  multiplication rules of the $A_4$ group. 

\vskip 1 cm
\section{Modular $A_4$ Group and Modular Forms of Level 3}

 The modular group $\overline{\G}$  
is the group of linear fractional transformations $\g$ 
acting on the complex variable $\t$ belonging to 
the upper-half complex plane as follows:
\be
\g\t = \frac{a\t + b}{c\t + d}\,,
\quad
\text{where} 
\quad
a,b,c,d \in \mathbb{Z}
\quad
\text{and}
\quad
ad - bc = 1\,,~~{\rm Im}\tau > 0\,.
\label{eq:linfractransform}
\ee
%
The group  $\overline{\G}$ is generated by two 
transformations $S$ and $T$ satisfying
%
\be
S^2 = \left(ST\right)^3 = I\,,
\ee
%
where $I$ is the identity element.
Representing $S$ and $T$ as
\be
S = \begin{pmatrix}
0 & 1 \\
-1 & 0
\end{pmatrix}\,,
\qquad
T = \begin{pmatrix} 
1 & 1 \\
0 & 1
\end{pmatrix}\,,
\ee
%
one finds
\be
\t \xrightarrow{S} -\frac{1}{\t}\,, 
\qquad
\t \xrightarrow{T} \t + 1\,.
\ee
%

The modular group $\overline{\G}$ 
is isomorphic to the projective special linear group 
$PSL(2,\mathbb{Z}) = SL(2,\mathbb{Z})/\mathbb{Z}_2$, 
where $SL(2,\mathbb{Z})$ is the special 
linear group of $2\times2$ matrices with integer elements 
and unit determinant, and $\mathbb{Z}_2 = \{I, -I\}$ is 
its centre ($I$ being the identity element).  
The special linear group $SL(2,\mathbb{Z}) = \G(1) \equiv \G$ contains a 
series of infinite normal subgroups $\G(N)$, $N = 1,2,3,\dots$:
\be
\G(N) = \left\{
\begin{pmatrix}
a & b \\
c & d
\end{pmatrix} 
\in SL(2,\mathbb{Z})\,, 
\quad
\begin{pmatrix}
a & b \\
c & d
\end{pmatrix} =
\begin{pmatrix}
1 & 0 \\
0 & 1
\end{pmatrix} 
~~(\text{mod } N)
\right\},
\ee
%
called the principal congruence subgroups.
For $N=1$ and $2$, we define the 
groups $\overline{\G}(N) \equiv \G(N)/\{I,-I\}$ with 
$\overline{\G}(1) \equiv \overline{\G}$).
For $N > 2$, $\overline{\G}(N) \equiv \G(N)$ 
since $\G(N)$ does not contain the subgroup $\{I,-I\}$.
For each $N$, the associated linear fractional transformations 
of the form in eq.~\eqref{eq:linfractransform} 
are in a one-to-one correspondence with 
the elements of $\overline{\G}(N)$. 

  The quotient groups $\G_N \equiv \overline{\G}/\overline{\G}(N)$ 
are called finite modular groups. 
For $N \leq 5$, these groups are isomorphic to 
non-Abelian discrete 
groups widely used in flavour model 
building (see, e.g., \cite{deAdelhartToorop:2011re}): 
$\G_2 \simeq S_3$, $\G_3 \simeq A_4$, $\G_4 \simeq S_4$ and 
$\G_5 \simeq A_5$. We will be interested in the 
finite modular group  $\G_3 \simeq A_4$.

 Modular forms of weight $k$ and level $N$ are holomorphic functions $f(\t)$ 
transforming under the action of $\overline{\G}(N)$ in the following way:
\be
f\left(\g\t\right) = \left(c\t + d\right)^k f(\t)\,, 
\quad 
\g \in \overline{\G}(N)\,.
\ee
%
Here $k$ is even and non-negative, and $N$ is natural.
Modular forms of weight $k$ and level $N$ span a linear space
of finite dimension.
The dimension of the linear space  
of modular forms of weight $k$ and level 3,
${\cal M}_k(\Gamma_3\simeq A_4)$, is $k+1$. 
There exists a basis in this space such that 
a multiplet of modular forms $f_i(\t)$ 
transforms according to
a unitary representation $\rho$ of the finite group $\G_N$:
\be
f_i\left(\g\t\right) = \left(c\t + d\right)^k \rho\left(\g\right)_{ij} f_j(\t)\,, 
\quad 
\g \in \overline{\G}\,.
\label{eq:vvmodforms}
\ee
%
In the case of $N=3$ of interest, the three linear independent  
weight 2 modular forms 
form a triplet of $A_4$ \cite{Feruglio:2017spp}. 
These forms have been explicitly obtained \cite{Feruglio:2017spp} in terms of
the Dedekind eta-function $\eta(\tau)$: 
\begin{equation}
\eta(\tau) = q^{1/24} \prod_{n =1}^\infty (1-q^n)~,
\end{equation}
%
where $q = e^{2 \pi i \tau}$. 
In what follows we will use the following basis of the 
$A_4$ generators  $S$ and $T$ in the triplet representation:
\begin{align}
\begin{aligned}
S=\frac{1}{3}
\begin{pmatrix}
-1 & 2 & 2 \\
2 &-1 & 2 \\
2 & 2 &-1
\end{pmatrix},
\end{aligned}
\qquad 
\begin{aligned}
T=
\begin{pmatrix}
1 & 0& 0 \\
0 &\omega& 0 \\
0 & 0 & \omega^2
\end{pmatrix}, 
\end{aligned}
\end{align}
%
where $\omega=e^{i\frac{2}{3}\pi}$ .
The  modular forms $(Y_1^{(2)},Y_2^{(2)},Y_3^{(2)})$ transforming
as a triplet of $A_4$ can be written in terms of 
$\eta(\tau)$ and its derivative \cite{Feruglio:2017spp}:
\begin{eqnarray} 
\label{eq:Y-A4}
Y_1^{(2)}(\tau) &=& \frac{i}{2\pi}\left( \frac{\eta'(\tau/3)}{\eta(\tau/3)}  +\frac{\eta'((\tau +1)/3)}{\eta((\tau+1)/3)}  
+\frac{\eta'((\tau +2)/3)}{\eta((\tau+2)/3)} - \frac{27\eta'(3\tau)}{\eta(3\tau)}  \right), \nonumber \\
Y_2^{(2)}(\tau) &=& \frac{-i}{\pi}\left( \frac{\eta'(\tau/3)}{\eta(\tau/3)}  +\omega^2\frac{\eta'((\tau +1)/3)}{\eta((\tau+1)/3)}  
+\omega \frac{\eta'((\tau +2)/3)}{\eta((\tau+2)/3)}  \right) , \label{tripletY} \\ 
Y_3^{(2)}(\tau) &=& \frac{-i}{\pi}\left( \frac{\eta'(\tau/3)}{\eta(\tau/3)}  +\omega\frac{\eta'((\tau +1)/3)}{\eta((\tau+1)/3)}  
+\omega^2 \frac{\eta'((\tau +2)/3)}{\eta((\tau+2)/3)}  \right)\,.
\nonumber
\end{eqnarray}
%
The overall coefficient in eq. (\ref{tripletY}) is 
one possible choice; it cannot be uniquely determined.
The triplet modular forms $Y_{1,2,3}^{(2)}$
have the following  $q$-expansions:
\begin{align}
Y^{(2)}=\begin{pmatrix}Y_1^{(2)}(\tau)\\Y_2^{(2)}(\tau)\\Y_3^{(2)}(\tau)\end{pmatrix}=
\begin{pmatrix}
1+12q+36q^2+12q^3+\dots \\
-6q^{1/3}(1+7q+8q^2+\dots) \\
-18q^{2/3}(1+2q+5q^2+\dots)\end{pmatrix}.
\end{align}
%
They satisfy also the constraint \cite{Feruglio:2017spp}:
\begin{align}
(Y^{(2)}_2)^2+2Y^{(2)}_1 Y^{(2)}_3=0~.
\label{condition}
\end{align}
%

%
\section{Residual Symmetries of $A_4$ and Modular Forms}
%

 Residual symmetries arise whenever the VEV of the modulus $\tau$ breaks
the modular group $\Gamma$ only partially,
i.e., the little group (stabiliser) of $\langle \tau \rangle$
is non-trivial. Residual symmetries have been investigated in 
the case of modular $S_4$ invariance in \cite{Novichkov:2018ovf}, 
and of $A_5$ invariance in \cite{Novichkov:2018xyzw},
where viable models of lepton masses and mixing have 
also been constructed. In the present work we consider 
models of lepton flavour based on the residual symmetries 
of the modular $A_4$ invariance.

 There are only 2 inequivalent finite points with 
non-trivial little groups of $\bar{\Gamma}$,
namely, $\langle \tau \rangle=-1/2+ i \sqrt{3}/2\equiv \tau_L$
and  $\langle \tau \rangle= i\equiv \tau_C$ \cite{Novichkov:2018ovf}.
The first point is the left cusp in the fundamental domain of the modular group,
which is invariant under the  $ST$ transformation $\tau=-1/(\tau+1)$.
Indeed, $\mathbb{Z}_3^{ST}=\{ I, ST,(ST)^2 \}$ is  one of  subgroups of 
$A_4$ group 
\footnote{In the recent publication \cite{deAnda:2018ecu} the 
authors obtain $\langle \tau \rangle=-1/2+ i \sqrt{3}/2\equiv \tau_L$ 
in a $SU(5)$ GUT theory with modular $A_4$ symmetry.}. 
The right cusp at  
$\langle \tau \rangle=1/2+ i \sqrt{3}/2\equiv \tau_R$
is related to $\tau_L$ by the $T$ transformation.
The $\langle \tau \rangle= i$ point is invariant under the $S$ transformation
$\tau=-1/\tau$. The subgroup $\mathbb{Z}_2^{S}=\{ I, S \}$ of $A_4$  
is preserved at $\langle \tau \rangle= \tau_C$.
There is also infinite point $\langle \tau \rangle= i \infty\equiv \tau_T$,
in which  the subgroup  $\mathbb{Z}^T_3=\{I,T,T^2 \}$ of $A_4$
is preserved.

 It is possible to calculate the values of the  $A_4$ triplet modular 
forms of weight 2 at the symmetry points $\tau_L$,  $\tau_C$ and  $\tau_T$.
The results are reported in Table~\ref{tb:modularforms}
in which the values of the modular forms at 
$\langle \tau \rangle =\tau_R$ are also 
given, to be compared with those at the other two points.

 As we have noted, the dimension of the linear space 
${\cal M}_k(\Gamma_3\simeq A_4)$ 
of modular forms of weight $k$ and level 3 is $k+1$. The modular forms of
weights higher than 2 can be obtained from the 
modular forms of weight 2. They transform according to certain 
irreducible representations of the $A_4$ group.
 Indeed,  for  weight 4 we have $5$  independent  modular forms,
 which are constructed by the weight $2$ modular forms through 
 the tensor product of  ${\bf 3\times 3}$ (see Appendix A).
  We obtain  one triplet ${\bf 3}$ and two singlets ${\bf 1}$,
  ${\bf 1'}$, while the third singlet ${\bf 1"}$ vanishes:
\begin{align}
\begin{aligned}
{\bf Y_3^{(4)}}\equiv 
\begin{pmatrix}
Y_1^{(4)}  \\
Y_2^{(4)} \\
Y_3^{(4)} 
\end{pmatrix}
=\frac{2}{3}
\begin{pmatrix}
(Y_1^{(2)} )^2 -Y_2^{(2)}Y_3^{(2)}  \\
(Y_3^{(2)} )^2 -Y_1^{(2)}Y_2^{(2)} \\
(Y_2^{(2)} )^2 -Y_1^{(2)}Y_3^{(2)} 
\end{pmatrix},
\end{aligned} \hskip 5 cm\\
\nonumber \\
{\bf Y_1^{(4)}}= ( Y^{(2)}_1)^2 +2Y^{(2)}_2 Y^{(2)}_3, \quad
{\bf Y_{1'}^{(4)}}= ( Y^{(2)}_3)^2 +2Y^{(2)}_1 Y^{(2)}_2, \quad
{\bf Y_{1"}^{(4)}}= ( Y^{(2)}_2)^2 +2Y^{(2)}_1 Y^{(2)}_3  \equiv 0\,
\label{Yj4}
\end{align}
%
where the vanishing ${\bf Y_{1"}^{(4)}}$ is due to the condition
in Eq. (\ref{condition}). 
Using Eq. (\ref{Yj4}) we can calculate the values of the modular 
forms of weight 4, transforming as ${\bf 3}$ and ${\bf \{ 1, ~ 1'\}}$,
at the symmetry points $\tau_L$,  $\tau_C$ and  $\tau_T$.
We show the results also in Table~\ref{tb:modularforms}.
\begin{table}[h]
	\centering
	\begin{tabular}{|c|c|c|c|} \hline 
		\rule[14pt]{0pt}{1pt}
		 & weight $2$ &weight $4$ &\\ 
		$\tau$ & \quad$\bf 3$ & $\bf 3$\hskip 3.5 cm $\{\bf 1,\quad 1'\}$& $Y_1^{(2)}$\\ \hline \hline 
		\rule[14pt]{0pt}{3pt}
		$\tau_L$	& $Y_1^{(2)}(1,\omega, -\frac{1}{2}\omega^2)$  &
		$3(Y_1^{(2)})^2(1, -\frac{1}{2}\omega, \omega^2)$, 
		\qquad $\{0,
		\quad \frac{9}{4}(Y_1^{(2)})^2 \omega\}$  &$0.9486...$\\ \hline
		\rule[14pt]{0pt}{3pt}
		$\tau_R$ 	& $Y_1^{(2)}(1,\omega^2, -\frac{1}{2}\omega)$ &
		$3(Y_1^{(2)})^2(1, -\frac{1}{2}\omega^2, \omega)$,
		\qquad $\{0,\quad \frac{9}{4}(Y_1^{(2)})^2 \omega^2\}$&$0.9486...$\\ \hline
		\rule[14pt]{0pt}{3pt}
		$\tau_C$ 	& $Y_1^{(2)}(1,1-\sqrt{3}, -2+\sqrt{3})$ &
		$(Y_1^{(2)})^2(1,1,1)$,\qquad $(Y_1^{(2)})^2\{6\sqrt{3}-9,
		\ \  9-6\sqrt{3} \}$ & $1.0225...$\\ \hline
		\rule[14pt]{0pt}{3pt}
		$\tau_T$ 	& $Y_1^{(2)}(1,0,0)$  &
		$(Y_1^{(2)})^2(1,0,0)$, \qquad $\{ (Y_1^{(2)})^2, \quad 0\}$ &$1$\\ \hline
	\end{tabular}
	\caption{Modular forms of weight $2$ and $4$ and the magnitude of 
		$Y_1^{(2)}$ at relevant $\tau$.
	}
	\label{tb:modularforms}
\end{table}

%
\section{Lepton Mass Matrices with Residual Symmetry}
%
%

We will consider next modular invariant lepton flavour models with the
$A_4$ symmetry, assuming that the massive neutrinos 
are Majorana particles and 
that the neutrino masses originate from the Weinberg 
dimension 5 operator.
There is a certain freedom for the assignments of irreducible representations 
and modular weights to leptons. We suppose that 
three left-handed (LH) lepton doublets form a triplet of the $A_4$ group.
The Higgs doublets are supposed to be zero weight singlets of $A_4$.
The generic assignments of representations and modular weights $k_I$  
to the MSSM fields
\footnote{For the modular weights of chiral superfields we follow the sign convention which is opposite to that of the modular forms, i.e. a field $\phi^{(I)}$ transforms as $\phi^{(I)} \to (c\tau+d)^{-k_I} \rho^{(I)}(\gamma)\, \phi^{(I)}$ under the modular transformation~$\gamma$.}
are presented in Table~\ref{tb:fields}. 
In order to construct  models with minimal number of parameters, 
we introduce no flavons.
For the charged leptons, we assign the three right-handed (RH) 
charged lepton fields for three different singlet representations  
of $A_4$, ${\bf( 1,1',1'')}$.
Therefore, there are three independent coupling constants 
in the superpotential of the charged lepton sector.
These coupling constants can be adjusted to the  
observed charged lepton masses.
Since there are three singlet irreducible representations 
in the $A_4$ group, there are six cases for the assignment of  
the three RH charged lepton fields.
However, this ambiguity does not affect the matrix 
which acts on the 
LH charged lepton fields and 
enters into the expression for the PMNS matrix.
Thus, effectively we have the following unique 
form for the superpotential:
\begin{align}
w_e&=\alpha e_RH_d(LY)_{\mathbf{1}}+\beta \mu_RH_d(LY)_{\mathbf{1'}}+\gamma \tau_RH_d(LY)_{\mathbf{1''}}~,
\label{charged} 
\end{align}
\begin{align}
w_\nu&=-\frac{1}{\Lambda}(H_u H_u LLY)_{\bf 1}~,
\label{Weinberg}
\end{align}
%
where the sums of the modular weights  should vanish.
The parameters $\alpha$, $\beta$, $\gamma$ and $\Lambda$
are constant coefficients.
\begin{table}[h]
	\centering
	\begin{tabular}{|c||c|c|c|c|c|} \hline
		&$L$&$(e_R,\mu_R,\tau_R)$&$H_u$&$H_d$&$Y$\\  \hline\hline 
		\rule[14pt]{0pt}{0pt}
		$SU(2)$&$\bf 2$&$\bf 1$&$\bf 2$&$\bf 2$&$\bf 1$\\
		$A_4$&$3$& \bf (1,\ 1'',\ 1')&$\bf 1$&$\bf 1$&$\bf 3,~ 1,~ 1'$\\
		$k_I$&$ k_L$&$(k_{e_R},k_{\mu_R},k_{\tau_R})$&0&0&$k$ \\ \hline
	\end{tabular}
	\caption{
		The charge assignment of $SU(2)$, $A_4$, and  modular weights ($k_I$ for fields and $k$ for coupling $Y$).
		The right-handed charged leptons are assigned three $A_4$ singlets, respectively.}
	\label{tb:fields}
\end{table}

%
\subsection{Charged Lepton Mass Matrix with Residual Symmetry}
%
%
By using the decomposition of the $A_4$ tensor products 
given in Appendix A, the superpotential in Eq. (\ref{charged}) leads  
to a mass matrix of charged leptons, which is written in terms of
modular forms of $A_4$ triplet with weight $k$:
\begin{align}
\begin{aligned}
M_E= v_d
\begin{pmatrix}
\alpha & 0 & 0 \\
0 &\beta & 0\\
0 & 0 &\gamma
\end{pmatrix}
\begin{pmatrix}
Y_1^{(k)} & Y_3^{(k)}& Y_2^{(k)} \\
Y_2^{(k)} & Y_1^{(k)} &  Y_3^{(k)} \\
Y_3^{(k)} &  Y_2^{(k)}&  Y_1^{(k)}
\end{pmatrix}_{RL}\, , 
\end{aligned}
\end{align}
%
where $v_d \equiv \vev{H_d^0}$.
Without loss of generality the  coefficients $\alpha$, $\beta$, and $\gamma$ 
can be made real positive by rephasing the  
RH charged lepton fields. 

We will discuss next the charged lepton mass matrix at the specific 
points of  $\tau=\tau_L,\tau_R~\tau_C,\tau_T$ in the case of weight $k=2$.
At $\tau=\tau_L$,  the matrix $M_E^\dagger M_E$, which is relevant for 
the left-handed mixing, is given as: 
 \begin{align}
\begin{aligned}
&M_E^\dagger M_E =\frac{9}{4} v_d^2\, (Y_1^{(2)})^2 \times \\
&\begin{pmatrix}
\alpha^2+\beta^2+\gamma^2/4 & -\omega^2/2 \alpha^2 + \omega^2 \beta^2 - \omega^2/2 \gamma^2 & \omega \alpha^2 - \omega/2 \beta^2 - \omega/2 \gamma^2 \\
-\omega/2 \alpha^2 + \omega \beta^2 - \omega/2 \gamma^2 & \alpha^2/4 + \beta^2 + \gamma^2 & -\omega^2/2 \alpha^2 -\omega^2/2 \beta^2 + \omega^2 \gamma^2 \\
\omega^2 \alpha^2 - \omega^2/2 \beta^2 - \omega^2/2 \gamma^2 & -\omega/2 \alpha^2 - \omega/2 \beta^2 + \omega \gamma^2 & \alpha^2 + \beta^2/4 + \gamma^2
\end{pmatrix}.
\end{aligned}
\label{MEtauL}
\end{align}
%
It is easily noticed that this matrix commutes with $ST$, which is guaranteed by
the residual symmetry  $\mathbb{Z}_3^{ST}$ at  $\tau=\tau_L$, where
\begin{equation}
\begin{aligned}
ST & = \frac{1}{3}
\begin{pmatrix}
-1 & 2\omega & 2\omega^2 \\
2 & -\omega & 2\omega^2 \\
2 & 2\omega & -\omega^2
\end{pmatrix}.
\end{aligned}
\end{equation}
%
Both matrices $M_E^\dagger M_E$ and $ST$ are diagonalized by the unitary matrix 
$U_E$:
\begin{equation}
\begin{aligned}
&U_E  \equiv TS = \frac{1}{3}
\begin{pmatrix}
-1 & 2 & 2 \\
2\omega & -\omega & 2\omega \\
2\omega^2 & 2\omega^2 & -\omega^2
\end{pmatrix}, \\
&U_E^{\dagger} ST U_E  = T = \text{diag}\, (1, \omega, \omega^2), \qquad
U_E^{\dagger}  M_E^{\dagger} M_E U_E  = \frac{9}{4} v_d^2\, (Y_1^{(2)})^2\, \text{diag}(\gamma^2, \alpha^2, \beta^2),
\end{aligned}
\label{UEtauL}
\end{equation}
%
where $U_E$ is independent of parameters $\alpha,\beta,\gamma$.

On the other hand, at $\tau=\tau_R$, we have:
\begin{align}
\begin{aligned}
&M_E^\dagger M_E =\frac{9}{4} v_d^2\, (Y_1^{(2)})^2 \times \\
&\begin{pmatrix}
\alpha^2+\beta^2+\gamma^2/4 & -\omega/2 \alpha^2 + \omega \beta^2 - \omega/2 \gamma^2 & \omega^2 \alpha^2 - \omega^2/2 \beta^2 - \omega^2/2 \gamma^2 \\
-\omega^2/2 \alpha^2 + \omega^2\beta^2 - \omega^2/2 \gamma^2 & \alpha^2/4 + \beta^2 + \gamma^2 & -\omega/2 \alpha^2 -\omega/2 \beta^2 + \omega \gamma^2 \\
\omega \alpha^2 - \omega/2 \beta^2 - \omega/2 \gamma^2 & -\omega^2/2 \alpha^2 - \omega^2/2 \beta^2 + \omega^2 \gamma^2 & \alpha^2 + \beta^2/4 + \gamma^2
\end{pmatrix}\,.
\end{aligned} 
\label{MEtauR}
\end{align}
%
The matrix $M_E^\dagger M_E$ in Eq. (\ref{MEtauR}) 
commutes with
\begin{align}
\begin{aligned}
TS=\frac{1}{3}
\begin{pmatrix}
-1 & 2 & 2 \\
2\omega & -\omega & 2\omega \\
2\omega^2 & 2\omega^2 & -\omega^2
\end{pmatrix}.
\end{aligned}
\end{align}
%
The fact that $M_E^\dagger M_E$ and $TS$ commute is a consequence of 
the residual symmetry  $\mathbb{Z}_3^{TS}$ at  $\tau=\tau_R$.
The matrix $M_E^\dagger M_E$ and $ST$ is  diagonalized by the unitary matrix: 
\begin{equation}
\begin{aligned}
U_E  \equiv ST = \frac{1}{3}
\begin{pmatrix}
-1 & 2\omega & 2\omega^2 \\
2 & -\omega & 2\omega^2 \\
2 & 2\omega & -\omega^2
\end{pmatrix}.
\end{aligned}
\label{UEtauR}
\end{equation}
%

At $\tau=\tau_C$,  the determinant of  $M_E$  vanishes.
Indeed, this mass matrix leads to a massless charged lepton,
and thus cannot be used for model building.

 Finally,  at $\tau=\tau_T$ we obtain the real diagonal matrix:
\begin{align}
\begin{aligned}
M_E = v_d\, Y_1^{(2)}
\begin{pmatrix}
\alpha & 0 & 0 \\
0 &\beta & 0\\
0 & 0 &\gamma
\end{pmatrix}.
\end{aligned}
\label{MEtauT}
\end{align}

In the case of modular forms of
weight 4 we can obtain a charged lepton mass matrix in which 
the modular forms transforming as ${\bf 1}$ and ${\bf 1'}$ 
do not contribute. As seen in Table~\ref{tb:modularforms},
the weight $4$ triplet modular forms 
coincide with weight $2$ ones at $\tau=\tau_L, \tau_R$.
Indeed, $M_E^\dagger M_E$ is obtained by replacing parameters
$(\alpha,\beta,\gamma)$ of the mass matrices  
in Eqs.~\eqref{MEtauL} and \eqref{MEtauR} 
with $(\gamma,\alpha,\beta)$, respectively.
Therefore, the mixing matrices in Eqs.~\eqref{UEtauL} 
and \eqref{UEtauR} are the same.

At $\tau=\tau_C$, the charged lepton mass matrix is of rank one, 
i.e., two massless charged leptons appear since the triplet modular 
forms are proportional to $(1,1,1)$. At $\tau=\tau_T$, 
the charged lepton mass matrix is equal to the diagonal one given 
in Eq.(\ref{MEtauT}) since the triplet weight 4 modular 
forms coincide with the weight $2$ modular forms.
 
%
\subsection{Neutrino Mass Matrix (Weinberg Operator) }
%
%

The neutrino mass matrix  is written in terms of $A_4$ triplet 
modular forms of weight $k$ by using the superpotential 
in Eq. (\ref{Weinberg}):
\begin{align}
\begin{aligned}
M_\nu= \frac{v_u^2}{\Lambda}
\begin{pmatrix}
2 Y_1^{(k)} & -Y_3^{(k)}& -Y_2^{(k)} \\
-Y_3^{(k)} & 2Y_2^{(k)} &  -Y_1^{(k)} \\
-Y_2^{(k)} &  -Y_1^{(k)}&  2 Y_3^{(k)}
\end{pmatrix}_{LL},
\end{aligned}
\end{align}
%
where $v_u \equiv \vev{H_u^0}$.

In the case of weight $2$ modular forms 
it is easily checked that  two lightest neutrino masses are degenerate 
at $\tau=\tau_L, \tau_R$, 
while the determinant of $M_\nu$ vanishes at $\tau=\tau_C$.
In the latter case one neutrino is massless and two neutrino 
masses are degenerate.
The two lightest neutrino masses are degenerate also at $\tau=\tau_T$.
It may be helpful to add a comment:
these degeneracies of neutrino masses still hold
even if we use the seesaw mechanism by introducing 
the three right-handed neutrino fields as $A_4$ triplet.
Thus, the realistic neutrino mass matrix is not obtained as far as we take  
weight $2$ modular forms at $\tau=\tau_L, \tau_R, \tau_C, \tau_T$.

In the case of weight $4$ modular forms, there is one candidate 
that can be consistent with the observed neutrino masses.
At $\tau=\tau_L, \tau_R$, the neutrino mass term  ${\bf 3_L 3_L Y^{(4)}_{3}}$ 
is similar to the case of  weight $2$, where two neutrino masses are degenerate.
 In the case of weight $4$, the singlet ${\bf 1'}$ also  contributes to 
the neutrino mass matrix through the coupling ${\bf 3_L 3_L Y^{(4)}_{1'}}$.  
However, this additional term cannot resolve the degeneracy.

It is easily noticed that two neutrino masses are degenerate also 
at $\tau=\tau_T$ since ${\bf Y_3^{(4)}}\sim (1, \ 0, \ 0)$. 
An additional $Y^{(4)}_{\bf 1}$ does not change this situation.

 At $\tau=\tau_C$, the triplet modular form, as seen in Table~\ref{tb:modularforms},
is ${\bf Y_3^{(4)}}\sim (1, \ 1, \ 1)$, which 
allows to get large mixing angles.
Moreover, we have ${\bf 1}$ and ${\bf 1'}$ modular functions.
Therefore, we expect  nearly tri-bimaximal mixing pattern of PMNS matrix
with three different massive neutrinos.
The LH weak-eigenstate neutrino fields couple
to  ${\bf Y_3^{(4)}}$. This coupling leads to 
the following neutrino Majorana mass matrix:
\begin{align}
\begin{aligned}
M_\nu= \frac{v_u^2}{\Lambda}\, (Y_1^{(2)})^2 
\begin{pmatrix}
2 & -1& -1 \\
-1 & 2 &  -1 \\
-1 &  -1&  2
\end{pmatrix}.
\end{aligned}
\label{triplet}
\end{align}
%
Moreover, the LH neutrino fields couple also to  
${\bf Y_1^{(4)}}$ and ${\bf Y_{1'}^{(4)}}$, 
which gives the following additional contributions to the neutrino 
Majorana mass matrix $M_\nu$:
\begin{align}
\begin{aligned}
3(2\sqrt{3}-3) \frac{v_u^2}{\Lambda}\, (Y_1^{(2)})^2 
\begin{pmatrix}
1 & 0&0 \\
0 & 0 &  1 \\
0&  1& 0
\end{pmatrix} \ , \qquad 
-3(2\sqrt{3}-3) \frac{v_u^2}{\Lambda}\, (Y_1^{(2)})^2 
\begin{pmatrix}
0 & 0&1 \\
0 & 1 &  0 \\
1&  0& 0
\end{pmatrix}\,,
\end{aligned}
\label{eq:singlet}
\end{align}
%
where each of these two terms enters $M_{\nu}$ with its own arbitrary constant.

To summarise, 
the charged lepton mass matrix could be consistent with observed masses
   at $\tau=\tau_L, \tau_R, \tau=\tau_T$ for both cases
    of weight $2$ and $4$ modular forms.
    On the other hand, the neutrino Majorana mass matrix is consistent with
    observed masses only at  $\tau=\tau_C$ for weight $4$ modular forms.
    There is no common symmetry value of $\tau$, 
which leads to charged lepton 
and neutrino masses 
that are consistent with the data.    
%
\section{Models with Residual Symmetry}
%

\vskip 0.2 cm
As seen in  the previous section, 
we could not find  models 
with one modulus $\tau$ and with residual symmetry,
which are phenomenologically viable.
Therefore, we consider the case having two moduli in the theory:
one $\tau^\ell$, responsible via its VEV for the breaking of the modular $A_4$
symmetry in the charged lepton sector, and the another one $\tau^\nu$,
breaking the modular symmetry in the neutrino sector. 
Our approach here is purely phenomenological. 	
Constructing a model with two different moduli
in the charged lepton and neutrino sectors is out of the scope of our study, 
it is a subject of ongoing research and work in progress.
However, there are hints from the recent study \cite{Baur:1901xyz}
that this might be possible 
\footnote{The authors of \cite{Baur:1901xyz} write in the Conclusions: 
`` As  we  find  different  flavor symmetries at different points in 
moduli space (in particular in six compact dimensions), fields
that  live  at  different  locations  in  moduli  space  feel  
a  different  amount  of  flavor  symmetry. 
(...)  This could lead to a different flavor- and CP -structure for the 
various sectors of the standard model like
up- or  down-quarks, charged  leptons  or  neutrinos.'' 
}. 
A model with two different moduli in the quark and lepton sectors, 
associated respectively with $S_3$ and $A_4$ modular symmetries, 
has been presented recently in \cite{Kobayashi:2018wkl}. 
 
  We present next our setup.
For the  charged lepton mass matrix,  
we take weight $2$ modular forms at  $\tau^\ell=\tau_T$ (Case I) 
or at $\tau^\ell=\tau_L$ (Case II)
\footnote{The same numerical  results are obtained at $\tau_R$ for
	weight $2$ modular forms. Weight $4$ modular forms lead also to
the same results at $\tau_L$ and $\tau_R$.}.
At the same time we 
use weight $4$ modular forms at $\tau^\nu=\tau_C$ 
for constructing the neutrino Majorana mass term.
In order for the modular weight in the superpotential to vanish,
we assign the following weights to the LH lepton and 
RH charged lepton fields:
\begin{equation}
k_L=2 \ , \qquad  k_{e_R}=k_{\mu_R}=k_{\tau_R}=0,
\label{weight}
\end{equation}
%
where the notations are self-explanatory.
We note that $k_L=2$ is common in both $\tau^\ell$ and
$\tau^\nu$ modular spaces.

Then, the charged lepton mass matrix is obtained 
by using as input the expressions for the weight $2$ 
modular forms given in Table~\ref{tb:modularforms}.
At  $\tau_T$, it is a diagonal matrix:
\begin{align}
\begin{aligned}
M_E = v_d
\begin{pmatrix}
\alpha & 0 & 0 \\
0 &\beta & 0\\
0 & 0 &\gamma
\end{pmatrix}\ : \quad \text{Case I}\,.
\label{eq:MeI}
\end{aligned}
\end{align}
%
At $\tau=\tau_L$, the charged lepton mass matrix has the form: 
\begin{align}
\begin{aligned}
M_E= v_d
\begin{pmatrix}
\alpha & 0 & 0 \\
0 &\beta & 0\\
0 & 0 &\gamma
\end{pmatrix}
\begin{pmatrix}
1 & \omega^2& -\frac{1}{2}\omega \\
-\frac{1}{2}\omega  & 1 &  \omega^2 \\
 \omega^2 &  -\frac{1}{2}\omega &  1
\end{pmatrix}_{RL} ~ : \quad \text{Case II} . 
\label{eq:MeII}
\end{aligned}
\end{align}
%
 The matrix $M_E^\dagger M_E$, which is relevant for the
calculation of the left-handed mixing, 
is given in Eq.~\eqref{MEtauL}.

 The neutrino mass matrix represents a sum of the contributions of
modular forms of ${\bf 3}$, ${\bf 1}$ and ${\bf 1'}$, 
with the terms involving the two singlet 
modular forms entering the sum with 
arbitrary  complex coefficients $A$ and $B$:
\begin{align}
\begin{aligned} M_\nu= \frac{v_u^2}{\Lambda}\, (Y_1^{(2)})^2 \left\{
\begin{pmatrix}
2  & -1& -1 \\
-1 & 2 &  -1 \\
-1&  -1&  2
\end{pmatrix}
+ \left [A
\begin{pmatrix}
1 & 0&0 \\
0 & 0 &  1 \\
0&  1& 0
\end{pmatrix} - B
\begin{pmatrix}
0 & 0&1 \\
0 & 1 &  0 \\
1&  0& 0
\end{pmatrix}\right ]\right\},
\end{aligned}
\label{eq:neutrinomass}
\end{align}
%
where the constants of the two terms in Eq.~\eqref{eq:singlet} 
are absorbed in the parameters $A$ and $B$.

The two models with charged lepton mass matrix $M_E$ 
specified in Eqs.~\eqref{eq:MeI} and \eqref{eq:MeII} and neutrino 
mass matrix $M_\nu$ given in Eq.~\eqref{eq:neutrinomass}, as we 
will show, lead to the same phenomenology. 
	
 As an alternative to the models with two moduli $\tau^\ell$ 
and  $\tau^\nu$, we present next a model with one modulus
$\tau^\nu$ and one flavon, breaking the modular symmetry 
to $\mathbb{Z}_2^S$ and $\mathbb{Z}_3^T$
in the neutrino and charged lepton sectors respectively
and leading to the charged lepton and neutrino mass matrices  
given in Eqs.~\eqref{eq:MeI} and \eqref{eq:neutrinomass}.
We introduce an $A_4$ triplet flavon $\phi$ with modular weight $k_{\phi} = -3$.
In contrast to Eq.~\eqref{weight}, the modular weights of 
the LH lepton doublet and 
RH charged lepton fields are chosen as follows:
\begin{equation}
k_L=2 \ , \qquad  k_{e_R}=k_{\mu_R}=k_{\tau_R}=1.
\end{equation}
%
As a consequence, the modular functions $Y^{(i)}$ do not couple to the 
charged lepton sector, but couple to the neutrino sector 
because $Y^{(i)}$ have  positive even modular weights.
On the other hand, the flavon $\phi$ couples only to the
charged lepton sector because of its odd weight 
\footnote{A similar construction in the charged lepton sector 
was presented in Ref.~\cite{Feruglio:2017spp}.}.
The corresponding terms of the superpotential are the same as given in
Eq.~\eqref{charged} with the modular form $Y$ replaced by the flavon
$\phi$.
Moreover, we can easily obtain the requisite VEV $\phi=v_E(1,0,0)^T$
preserving $\mathbb{Z}_3^T$, $v_E$ being a constant parameter, 
from the potential analysis as seen in 
Refs.~\cite{Altarelli:2005yp,Altarelli:2005yx}.
Finally, we get the charged lepton and neutrino mass matrices  
given in Eqs.~\eqref{eq:MeI} and \eqref{eq:neutrinomass}.
This flavon model with one modulus $\tau^\nu$
leads to the same phenomenology as the models considered earlier 
with two different moduli  $\tau^\ell$ and  $\tau^\nu$.  

%
\subsection{The Neutrino Mixing}
%

 In case I, only the neutrino mass matrix contributes to 
the PMNS matrix since the charged lepton mass matrix is diagonal.
The neutrino mass matrix in this case leads 
to the so called $\text{TM}_2$ mixing form  
of PMNS matrix $U_{\text{PMNS}}$ \cite{Grimus:2008tt,Albright:2010ap} 
where the second column of $U_{\text{PMNS}}$ is trimaximal:
\begin{equation}
U_{\text{PMNS}}^{\,\text{I}}=
\begin{pmatrix}
\frac{2}{\sqrt{6}} & \frac{1}{\sqrt{3}} & 0 \\
-\frac{1}{\sqrt{6}} & \frac{1}{\sqrt{3}} & -\frac{1}{\sqrt{2}} \\
-\frac{1}{\sqrt{6}} & \frac{1}{\sqrt{3}} & \frac{1}{\sqrt{2}}
\end{pmatrix}
\begin{pmatrix}
\cos \theta & 0 & e^{i\phi }\sin \theta \\
0 & 1 & 0 \\
-e^{-i\phi }\sin \theta & 0 & \cos \theta 
\end{pmatrix}\,{\rm P}\,.
\label{eq:TM2}
\end{equation}
%
Here $\theta$ and $\phi$ are arbitrary mixing angle and phase, 
respectively, and $P$ is a diagonal phase matrix containing 
contributions to the Majorana phases of $U_{\text{PMNS}}$.
Employing the standard parametrisation of  $U_{\text{PMNS}}$ 
(see, e.g., \cite{Tanabashi:2018oca}), 
it is possible to show that 
the trimaximal mixing pattern
leads to the following relation between 
the reactor angle $\theta_{13}$ and $\theta$, 
between the atmospheric neutrino 
mixing angle $\theta_{23}$ and $\theta_{13}$ and $\theta$,
and sum rules for the solar neutrino
mixing angle $\theta_{12}$ 
and for the Dirac phase $\delta$
 \cite{Grimus:2008tt,Albright:2010ap} 
(see also \cite{Petcov:2017ggy,Girardi:2015rwa}):
\begin{align}
  \label{eq:theta13_theta}
  \sin^2 \theta_{13} &= \frac{2}{3} \sin^2 \theta\,,\\
\label{eq:theta12_theta13}
  \sin^2 \theta_{12}  &= \frac{1}{3\,\cos^2\theta_{13}}\,, \\
\label{eq:theta23_phi}
  \sin^2 \theta_{23} &= \frac{1}{2} + \frac{s_{13}}{2} \frac{\sqrt{2 - 3 s_{13}^2}}{1 - s_{13}^2} \cos \phi\,, \\
\label{eq:delta_thetas}
\cos\delta &= \frac{\cos2\theta_{23}\,\cos2\theta_{13}}
{\sin2\theta_{23}\,\sin\theta_{13}\,(2 - 3\sin^2\theta_{13})^\frac{1}{2}}\,.
\end{align}
%

Using the $3\sigma$ allowed range of $\sin^2\theta_{13}$ 
from \cite{Capozzi:2018ubv} 
and Eq.~\eqref{eq:theta13_theta} 
we get the following constraints on $\sin\theta$:
\begin{equation}
  0.17 \lesssim |\sin \theta| \lesssim 0.19\,.
\end{equation}
%
To leading order in $s_{13}$ we obtain from Eq.~\eqref{eq:theta23_phi}:
\be 
 \dfrac{1}{2} -  \dfrac{s_{13}}{\sqrt{2}}
\ltap  \sin^2\theta_{23} \ltap 
\dfrac{1}{2} +  \dfrac{s_{13}}{\sqrt{2}}  
\,,~~{\rm or}~~
 0.391~(0.390)\ltap  \sin^2\theta_{23} \ltap 0.609~(0.611)\,,
\label{A4s2232}
\ee
%
where the numerical values correspond to 
the maximal allowed value of $\sin^2\theta_{13}$
at $3\sigma$ C.L. for NO (IO) neutrino mass spectrum 
\cite{Capozzi:2018ubv}.
The interval of possible values of  
$\sin^2\theta_{23}$ in eq. (\ref{A4s2232}) 
is somewhat wider that the $3\sigma$ ranges of 
experimentally allowed values of $\sin^2\theta_{23}$ for NO 
and IO spectra given in \cite{Capozzi:2018ubv}.
Using the $3\sigma$ allowed ranges of  $\sin^2\theta_{23}$
and $\sin^2\theta_{13}$ for NO (IO) spectra from \cite{Capozzi:2018ubv} 
and Eq. (\ref{eq:theta23_phi}) we also get:
\begin{equation}
 -\,0.640~(-\,0.508)\lesssim \cos \phi \leq 1\,. 
\label{eq:phirange}
\end{equation}
%

 The phase $\phi$ is related to the Dirac phase $\delta$ 
\cite{Petcov:2017ggy}:
\begin{equation}
\label{eq:deltaphi}
  \sin 2\theta_{23} \sin \delta = \sin \phi\,.
\end{equation}
%
The  Majorana phase $\alpha_{31}/2$ of 
the standard parametrisation of 
$U_{\text{PMNS}}$ \cite{Tanabashi:2018oca}
receives contributions from 
the phase $\phi$ via \cite{Petcov:2017ggy}
\be
\dfrac{\alpha_{31}}{2} = \dfrac{\xi_{31}}{2} + \alpha_{2} + \alpha_3\,,
\label{alpha31A4}
\ee
%
where the phase $\xi_{31}$ will be specified later, 
\be
\label{alpha2alpha3}
\alpha_2 = {\rm arg}
\big(-\,\dfrac{c}{\sqrt{2}} - \dfrac{s}{\sqrt{6}}\,e^{i\phi}\big)\,,~~~
\alpha_3 = {\rm arg}
\big(\dfrac{c}{\sqrt{2}} - \dfrac{s}{\sqrt{6}}\,e^{i\phi}\big)\,,
\ee
%
\begin{eqnarray}
\label{alpha2}
\sin\alpha_2 = -\,\dfrac{s}{\sqrt{6}}\,\dfrac{\sin\phi}{s_{23}\,c_{13}} = 
 -\,\tan\theta_{13}\,\cos\theta_{23}\,\sin\delta\,,\\[0.30cm]
\label{alpha3}
\sin\alpha_3 = -\,\dfrac{s}{\sqrt{6}}\,\dfrac{\sin\phi}{c_{23}\,c_{13}} = 
 -\,\tan\theta_{13}\,\sin\theta_{23}\,\sin\delta\,.
\end{eqnarray}
%
We also have \cite{Petcov:2017ggy}:
\be
\sin(\phi - \alpha_2 - \alpha_3) = -\,\sin\delta\,.
\label{sindsinalphab2b3A4}
\ee
%

 For further discussion of phenomenology of the
neutrino trimaximal mixing (\ref{eq:TM2}), see, e.g., 
\cite{Shimizu:2011xg,Girardi:2016zwz,Petcov:2017ggy,Shimizu:2014ria}.

 In  case II, the contribution of the rotation of the charged lepton sector 
is added to the trimaximal mixing,  which is derived from the 
neutrino mass matrix in Eq.~\eqref{eq:neutrinomass}.
The mixing matrix in the charged lepton sector is the matrix $U_E$ 
in Eq. (\ref{UEtauL}). The PMNS matrix is given by:
\begin{equation}
\begin{aligned}
U_{\text{PMNS}}^{\, \text{II}}=
 \frac{1}{3}
\begin{pmatrix}
-1 & 2 & 2 \\
2\omega & -\omega & 2\omega \\
2\omega^2 & 2\omega^2 & -\omega^2
\end{pmatrix}^\dagger
\begin{pmatrix}
\frac{2}{\sqrt{6}} & \frac{1}{\sqrt{3}} & 0 \\
-\frac{1}{\sqrt{6}} & \frac{1}{\sqrt{3}} & -\frac{1}{\sqrt{2}} \\
-\frac{1}{\sqrt{6}} & \frac{1}{\sqrt{3}} & \frac{1}{\sqrt{2}}
\end{pmatrix}
\begin{pmatrix}
\cos \theta & 0 & e^{i\phi }\sin \theta \\
0 & 1 & 0 \\
-e^{-i\phi }\sin \theta & 0 & \cos \theta 
\end{pmatrix}\,P.
\end{aligned}
\label{eq:PMNS2}
\end{equation}
%
It is straightforward to check that after a substitution 
\(\theta \to \theta - \pi/2\), \(\phi \to -\phi\), 
the PMNS matrix~\eqref{eq:PMNS2} can be rewritten as
\begin{equation}
  U_{\text{PMNS}}^{\, \text{II}} =
  \begin{pmatrix}
    -1 & 0 & 0 \\
    0 & e^{i \pi / 3} & 0 \\
    0 & 0 & e^{-i \pi / 3}
  \end{pmatrix}
  U_{\text{PMNS}}^{\, \text{I}}
  \begin{pmatrix}
    e^{i (\phi - \pi / 2)} & 0 & 0 \\
    0 & 1 & 0 \\
    0 & 0 & e^{-i (\phi + \pi / 2)}
  \end{pmatrix}.
  \label{eq:PMNS21corr}
\end{equation}
%
The leftmost phase matrix does not contribute to the mixing, 
since its effect can be absorbed into the charged lepton field phases.
The rightmost phase matrix contributes only to the Majorana phases, 
therefore the numerical predictions in this case are the same 
as in Case I,
apart possibly from the corresponding shift of the Majorana phases.
However, as can be shown analytically, and we have confirmed numerically, 
also the predictions for the Majorana phases in Case II coincide 
with the predictions in case I.

%
\subsection{The Neutrino Masses and Majorana Phases}
%

It follows from \eqref{eq:neutrinomass} that the neutrino mass 
matrix \(M_{\nu}\) is a linear combination of three basis matrices:
\begin{equation}
  M_1 =
  \begin{pmatrix}
    2 & -1 & -1 \\
    -1 & 2 & -1 \\
    -1 & -1 & 2
  \end{pmatrix},
  \quad
  M_2 =
  \begin{pmatrix}
    1 & 0 & 0 \\
    0 & 0 & 1 \\
    0 & 1 & 0
  \end{pmatrix},
  \quad
  M_3 =
  \begin{pmatrix}
    0 & 0 & 1 \\
    0 & 1 & 0 \\
    1 & 0 & 0
  \end{pmatrix}.
\end{equation}
%
To diagonalize \(M_{\nu}\), it is convenient to rewrite it in 
a different basis:
\begin{equation}
  \begin{aligned}
    M'_1 &= \frac{1}{\sqrt{3}} \left( M_2 + 2 M_3 \right) =
    \frac{1}{\sqrt{3}}
    \begin{pmatrix}
      1 & 0 & 2 \\
      0 & 2 & 1 \\
      2 & 1 & 0
    \end{pmatrix}, \\
    M'_2 &= M_2 + \frac{1}{3} M_1 =
    \frac{1}{3}
    \begin{pmatrix}
      5 & -1 & -1 \\
      -1 & 2 & 2 \\
      -1 & 2 & 2
    \end{pmatrix}, \\
    M'_3 &= M_2 - \frac{1}{3} M_1 =
    \frac{1}{3}
    \begin{pmatrix}
      1 & 1 & 1 \\
      1 & -2 & 4 \\
      1 & 4 & -2
    \end{pmatrix}, \\
    M_{\nu} &= c \left( M'_1 + a M'_2 + b M'_3 \right),
  \end{aligned}
\end{equation}
%
where \(a\) and \(b\) are arbitrary complex coefficients and \(c\) is 
the overall scale factor which can be rendered real positive.
\(M_{\nu}\) is diagonalized by a unitary matrix \(U_{\nu}^{\circ}\) 
of the following form:
\begin{equation}
  U_{\nu}^{\circ} = V_{\text{TBM}} \, U_{13} (\theta,\phi)\,,
\end{equation}
%
so that $M_{\nu} = (U^\circ_{\nu})^*\,M_{\nu}^{\text{diag}}\,(U_{\nu}^{\circ})^\dagger$,
with $M_{\nu}^{\text{diag}} = 
\diag \left( m_1 e^{-i 2\phi_1}, m_2 e^{-i 2\phi_2}, m_3 e^{-i 2\phi_3} \right)$, 
where $m_i e^{-i 2\phi_i}$ are complex eigenvalues and $m_i \geq 0$ are the neutrino masses.
\footnote{
In general, the standard labelling of the neutrino masses~\cite{Tanabashi:2018oca} corresponds to some permutation of the neutrino mass matrix eigenvalues, which affects the order of the PMNS matrix columns.
However, the only non-trivial permutation of the \(\text{TM}_2\) matrix columns consistent with the experimental data is \((321)\), which is equivalent to a shift \(\theta \to \theta - \pi/2\) up to an unphysical overall column sign.
Hence, we can assume that the order of neutrino mass matrix eigenvalues coincides with the standard labelling without loss of generality.
}
Extracting the phases $\phi_i$ from $M_{\nu}^{\text{diag}}$, we find:
\begin{equation}
 M_{\nu}^{\text{diag}} = e^{-i2\phi_1}\,P^*\,
\diag \left(m_1, m_2, m_3\right)\,P^*\,,
~P = \diag \left(1,e^{i(\phi_2-\phi_1)},e^{i(\phi_3-\phi_1)}\right)\,,
\end{equation}
%
where the phases $(\phi_2-\phi_1)$ and  $(\phi_3-\phi_1)$ 
contribute to the Majorana phases $\alpha_{21}/2$ and 
$\alpha_{31}/2$ of the standard parametrisation of the PMNS 
matrix  \cite{Tanabashi:2018oca}. 
Thus, the PMNS matrix has the form:
\begin{equation}
U_{\text{PMNS}} =  U_{\nu}^{\circ}\,P = 
 e^{-i2\phi_1}\,V_{\text{TBM}} \, U_{13} (\theta,\phi)\,P\,,
\end{equation}
%
where the common phase factor $e^{-i2\phi_1}$
is unphysical. The phase $\xi_{31}/2$ in 
Eq. (\ref{alpha31A4}) can be identified now with  
$(\phi_3-\phi_1)$: $\xi_{31}/2 = \phi_3-\phi_1$.
Thus, the Majorana phases $\alpha_{21}/2$ 
and $\alpha_{31}/2$ are given by:
\begin{equation}
\dfrac{\alpha_{21}}{2} = \phi_2 - \phi_1\,,~~~
\dfrac{\alpha_{31}}{2} = \phi_3 - \phi_1 + \alpha_2 + \alpha_3\,.
\label{eq:alpha2131}
\end{equation}
%

 The complex rotation parameters \(\theta\) and \(\phi\) 
are fixed by a choice of \(a\) and \(b\), which we will 
now show explicitly. We find by direct calculation that
\begin{equation}
  \begin{aligned}
    U_{\nu}^{\circ \, T} M'_1 \, U_{\nu}^{\circ} &=
    \begin{pmatrix}
      - e^{-i\phi} \sin 2\theta & 0 & \cos 2\theta \\
      0 & \sqrt{3} & 0 \\
      \cos 2\theta & 0 & e^{i\phi} \sin 2\theta
    \end{pmatrix}, \\
    U_{\nu}^{\circ \, T} M'_2 \, U_{\nu}^{\circ} &=
    \begin{pmatrix}
      2 \cos^2 \theta & 0 & e^{i\phi} \sin 2\theta \\
      0 & 1 & 0 \\
      e^{i\phi} \sin 2\theta & 0 & 2 e^{2i\phi} \sin^2 \theta
    \end{pmatrix}, \\
    U_{\nu}^{\circ \, T} M'_3 \, U_{\nu}^{\circ} &=
    \begin{pmatrix}
      -2 e^{-2i\phi} \sin^2 \theta & 0 & e^{-i\phi} \sin 2\theta \\
      0 & 1 & 0 \\
      e^{-i\phi} \sin 2\theta & 0 & -2 \cos^2 \theta
    \end{pmatrix}\,.
  \end{aligned}
  \label{eq:diagBasis}
\end{equation}
%
Thus, the neutrino mass matrix \(M_{\nu}\) is diagonalized when 
the corresponding linear combination of the off-diagonal entries vanishes, 
which leads to
\begin{equation}
  \cos 2\theta + a e^{i\phi} \sin 2\theta + b e^{-i\phi} \sin 2\theta = 0
  \quad
  \Leftrightarrow
  \quad
  a e^{i\phi} + b e^{-i\phi} = - \cot 2\theta.
  \label{eq:theta_abphi}
\end{equation}
%
The above condition is equivalent to:
\begin{equation}
  e^{i\phi} = \pm \frac{a^{*} - b}{\left| a^{*} - b \right|}, \quad
  \cot 2\theta = \mp \frac{|a|^2 - |b|^2}{\left| a^{*} - b \right|} \,.
  \label{eq:phitheta_ab}
\end{equation}
%

It proves convenient to introduce the complex parameter
\begin{equation}
  z = a e^{i\phi} - b e^{-i\phi}
  = \pm \frac{|a|^2 + |b|^2 - 2ab}{\left| a^{*} - b \right|}\,.
  \label{eq:z_ab}
\end{equation}
%
Using \((\theta, \phi, z)\) is a reparametrisation of \((a, b)\) 
determined by \eqref{eq:phitheta_ab} and \eqref{eq:z_ab}.
The inverse parameter transformation is given by
\begin{equation}
  \begin{aligned}
    a &= \frac{e^{-i\phi}}{2} \left( z - \cot 2\theta \right), \\
    b &= \frac{e^{i\phi}}{2} \left( -z - \cot 2\theta \right).
  \end{aligned}
\end{equation}
%

 The neutrino mass matrix eigenvalues are the corresponding linear 
combinations of the diagonal entries in \eqref{eq:diagBasis}:
\begin{equation}
  \begin{aligned}
    m_1 e^{-i(2\phi_1 - \phi)} &= c \left( z - \frac{1}{\sin 2\theta} \right), \\
    m_2 e^{-i 2 \phi_2} &= c \left( \sqrt{3} - i z \sin \phi - \cot 2\theta \cos \phi \right), \\
    m_3 e^{-i(2\phi_3 + \phi)} &= c \left( z + \frac{1}{\sin 2\theta} \right).
  \end{aligned}
\end{equation}
%

Fitting the mass-squared differences to experimentally observed values, we find the following constraint on \(z\) in terms of \(\theta\), \(\phi\) and \(r \equiv \Delta m_{21}^2 / \Delta m_{31}^2\):
\begin{equation}
  \left| z - z_0 \right|^2 = R^2,
  \quad \text{sign} \left( \re z \right) = \pm\, \text{sign} \left( \sin 2\theta \right),
  \label{eq:z_circle}
\end{equation}
%
where the plus (minus) sign corresponds to NO (IO) spectrum 
of neutrino masses, and
\begin{equation}
  \begin{aligned}
    z_0(\theta, \phi, r) &= \frac{1-2r}{\cos^2 \phi \sin 2\theta} + \tan \phi \left( \frac{\sqrt{3}}{\cos \phi} - \cot 2\theta \right) i, \\
    R^2(\theta, \phi, r) &= \left[ \left( \sqrt{3} - \cot 2\theta \cos \phi \right)^2 + \frac{(1-2r)^2 - \cos^2\phi}{\sin^2 2\theta} \right] \Big/ \cos^4 \phi.
  \end{aligned}
\end{equation}
%
Since \(\theta\) and \(r\) are tightly constrained by the experimental data, 
the set of phenomenologically viable models is effectively described by 
two angles \(\phi\) and \(\psi\), with the latter being the angle parameter 
on the circle~\eqref{eq:z_circle}, i.e. \(z = z_0 + R\, e^{i\psi}\).
Scanning through \(\phi\) and \(\psi\) numerically, we find that to each 
set of the experimentally allowed values of the mixing angles and 
the mass-squared differences corresponds a range of models 
(parameterised by \(\psi\)) with different values of the neutrino 
masses and the Majorana phases.

\begin{figure}[t!]
  \centering
  \includegraphics[width=\textwidth]{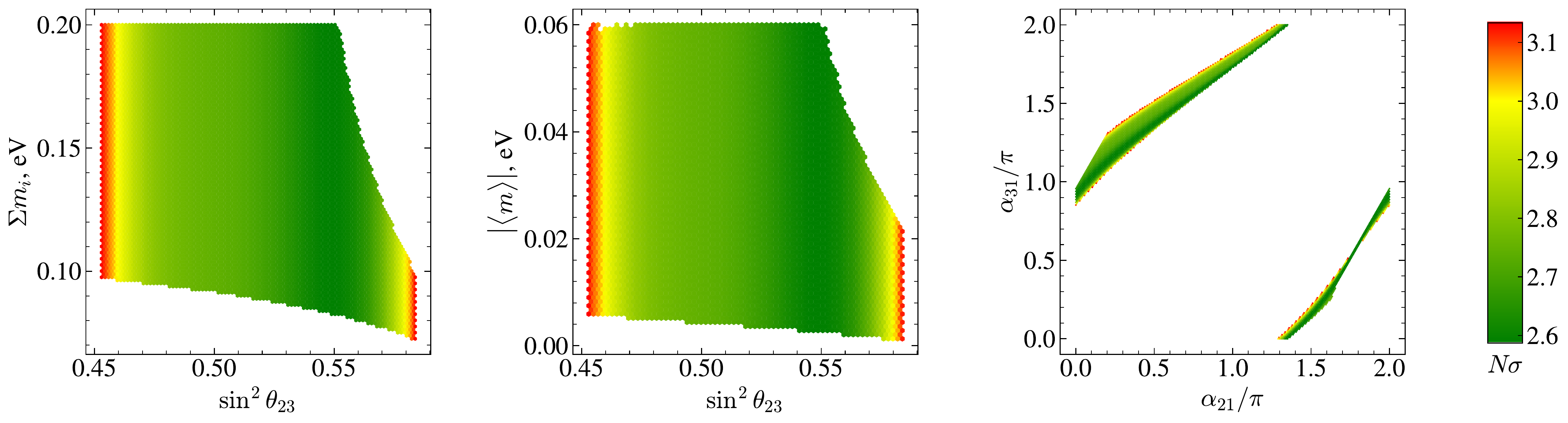}
  \caption{Correlations between \(\sin^2 \theta_{23}\) and 
the sum of neutrino masses \(\sum m_i\),
between \(\sin^2 \theta_{23}\) and the 
effective Majorana mass \(\left| \left\langle m \right\rangle \right|\),
and between the Majorana phases \(\alpha_{31}\) and \(\alpha_{21}\) 
in the case of NO neutrino mass spectrum. See text for further details.
  }
  \label{fig:corr}
\end{figure}

We report the numerical results in the case of NO spectrum 
in Fig.~\ref{fig:corr}.
The allowed range of the sum of neutrino masses depends on the 
value of $\sin^2 \theta_{23}$.
The lower bound slightly decreases from 0.097 eV to 0.074 eV as 
$\sin^2 \theta_{23}$ runs through its \(3\sigma\) confidence interval
of \([0.46, 0.58]\).
\footnote{We define the number of standard deviations from the \(\chi^2\) minimum as \(N\sigma = \sqrt{\Delta \chi^2}\), where \(\Delta \chi^2\) is a sum of one-dimensional projections \(\Delta \chi_j^2\), \(j = 1,2,3,4\) from~\cite{Capozzi:2018ubv} for the accurately known dimensionless observables \(\sin^2 \theta_{12}\), \(\sin^2 \theta_{13}\), \(\sin^2 \theta_{23}\) and \(r\).}
On the other hand, the upper bound is highly dependent on the value 
of \(\sin^2 \theta_{23}\),
and tends to infinity as \(\sin^2 \theta_{23}\) approaches 0.5, 
which corresponds to \(\delta = \phi = 3 \pi/2\).
This means that at this point the sum of neutrino masses is allowed 
to take any value greater than its lower bound of 0.093 eV.
The dependence of the effective Majorana mass 
\(\left| \left\langle m \right\rangle \right|\) on \(\sin^2 \theta_{23}\) 
is qualitatively similar to that of the sum of neutrino masses.
The maximal value of  
$\left| \left\langle m \right\rangle \right| \cong 0.059$ eV 
is practically independent of \(\sin^2 \theta_{23}\) 
for $0.46 \leq \sin^2 \theta_{23}\leq 0.55$. 
The lower bound of \(\left| \left\langle m \right\rangle \right|\) varies 
from 0.0015 eV to 0.0059 eV for \(\sin^2 \theta_{23}\) in its \(3\sigma\) range.
However, for values of $\sin^2 \theta_{23}$ from its $3\sigma$ range,  
$0.46 \leq \sin^2 \theta_{23}\leq 0.58$, 
$\left| \left\langle m \right\rangle \right|$ can  
have values in the interval $[0.0059,0.059]$ eV (see Fig. \ref{fig:corr}). 
Most (if not all) of these values  
may be probed in the future neutrinoless double beta decay experiments.

There is also a strong correlation between the Majorana phases.
The set of best-fit models corresponds to \(\phi = 1.664\pi\) and 
leads to the following values of observables:
\begin{equation}
  \begin{gathered}
    r = 0.0299,\quad
    \delta m^2 = 7.34 \cdot 10^{-5} \text{ eV}^2,\quad
    \Delta m^2 = 2.455 \cdot 10^{-3} \text{ eV}^2,\\
    \sin^2 \theta_{12} = 0.3406,\quad
    \sin^2 \theta_{13} = 0.02125,\quad
    \sin^2 \theta_{23} = 0.5511,\\
    m_1 = 0.0143-0.0612 \text{ eV},\quad
    m_2 = 0.0166-0.0618 \text{ eV},\quad
    m_3 = 0.0519-0.079 \text{ eV},\\
    \textstyle\sum_i m_i = 0.0828-0.2019 \text{ eV},\quad
    \left| \left\langle m \right\rangle \right| = 0.0029-0.0589 \text{ eV},\quad
    \delta / \pi = 1.339,
  \end{gathered}
  \label{eq:viableObs}
\end{equation}
%
consistent with the experimental data at 2.59$\sigma$ C.L..

Similar analysis can be performed in the case of IO neutrino mass spectrum.
However, in that case the minimal value of the sum of the three neutrino masses is 0.63 eV, and we do not analyse this case further.

%
\section{Summary}
%
%

We have investigated 
models of lepton masses and mixing 
based on modular $A_4$ 
flavour symmetry broken to residual
symmetries in the charged lepton and neutrino sectors.
The standard case of three lepton families was considered.
In a theory based on finite modular flavour symmetry 
not only the matter fields, but also
the constants such as the Yukawa couplings 
transform non-trivially under the modular symmetry.
These constants  are written in terms of modular forms which are  
holomorphic functions of a complex scalar field - the modulus  $\tau$. 
The modular forms have specific transformation properties 
under the modular symmetry transformations, which 
are characterised by a positive even number $k$ called ``weight'', 
and depend on the order of the finite modular group 
via their ``level'' $N$. 
In the case of  modular $A_4$ symmetry we have $N=3$ 
and for the lowest weight modular forms $k=2$.
The modular forms transform under the usual 
non-Abelian discrete flavor symmetries as well.
Modular forms of weight $k$ and level $N$ 
span a linear space of finite dimension.
There exists a basis in this space such that 
the modular forms form multiplets transforming 
according to unitary irreducible representations  
of the finite modular group. In the case of 
modular $A_4$ symmetry, the dimension of the 
linear space of modular forms of weight $k=2$ is 3,
and one can employ modular forms transforming 
as the triplet irreducible representation 
of  $A_4$. Modular forms of higher weights can be 
obtained as direct products of the modular forms 
of weight $k=2$.

  In lepton flavour models with finite modular symmetry,
the modular symmetry must be broken in order  
to distinguish between the electron, muon and tauon, 
generate three different neutrino masses and reproduce 
the measured values of the three neutrino mixing angles.
In the most economical versions of the flavour models 
the only source of breaking of the modular 
symmetry is the VEV of the modulus $\tau$,  
$\langle \tau \rangle \neq 0$, 
and there is no need to introduce flavon fields. 
In the present article 
we consider both a model
without flavons, in which the $A_4$ symmetry 
is broken only by $\langle \tau \rangle$, 
and a model with one triplet flavon field,
in which the $A_4$ symmetry is broken by $\langle \tau \rangle$
in the neutrino sector and by the VEV of the flavon in the 
charged lepton sector.

 The modular group $A_4$ has two generators
$S$ and $T$ satisfying the presentation rules:
$S^2 = (ST)^3 = T^3 = I$, where $I$ is the 
unit operator. Residual symmetries arise whenever the VEV of the 
modulus $\tau$ breaks the considered finite modular group 
$\Gamma_N$, $\Gamma_3 \simeq A_4$,
only partially, i.e., the little group (stabiliser) 
of $\langle \tau \rangle$ is non-trivial.
 There are only 2 inequivalent finite points with non-trivial little groups,
namely, $\langle \tau \rangle=-1/2+ i \sqrt{3}/2\equiv \tau_L$
and  $\langle \tau \rangle= i\equiv \tau_C$ \cite{Novichkov:2018ovf}.
The first one is the left cusp in the fundamental domain of the modular group,
and corresponds to a residual symmetry associated with the 
subgroup  $\mathbb{Z}_3^{ST}=\{ I, ST,(ST)^2 \}$ of the $A_4$ group.
The $\langle \tau \rangle= i$ point is invariant under 
the $S$ transformation ($\tau=-1/\tau$) 
of the  $\mathbb{Z}_2^{S}=\{ I, S \}$ subgroup of $A_4$.
There is also infinite point $\langle \tau \rangle= i \infty\equiv \tau_T$,
in which  the subgroup  $\mathbb{Z}^T_3=\{I,T,T^2 \}$ of $A_4$
is preserved.

 We have constructed phenomenologically viable models of lepton masses and 
mixing based on modular $A_4$ invariance broken to residual 
symmetries $\mathbb{Z}^{T}_3$ or $\mathbb{Z}^{ST}_3$ 
and $\mathbb{Z}^S_2$ respectively in the charged 
lepton and neutrino sectors. 
The neutrino Majorana mass term is assumed to be generated 
by the dimension 5 Weinberg operator.
We found that there is no common symmetry value of $\tau$, 
which leads to charged lepton and neutrino masses that 
are consistent with the data.   
For the  construction of the charged lepton mass matrix,  
we used weight $2$ modular forms at  $\tau^\ell=\tau_T$ (Case I) 
or at $\tau^\ell=\tau_L$ (Case II).
At the same time we used weight $4$ modular forms at $\tau^\nu=\tau_C$ 
for constructing the neutrino Majorana mass term. 
Since at present we are not aware of a mechanism 
that can lead to different values of $\langle \tau \rangle$ 
in the neutrino and the charged lepton sectors,
our analysis only suggests  
that having trimaximal neutrino mixing in models 
with modular $A_4$ symmetry without flavons might be possible.
We also show that, alternatively, trimaximal neutrino mixing 
can be obtained by assuming that $\langle \tau \rangle$ breaks 
the $A_4$ symmetry to $\mathbb{Z}^S_2$ in the neutrino sector, while 
a VEV of a triplet flavon field 
breaks the $A_4$ symmetry to $\mathbb{Z}^{T}_3$
in the charged lepton sector.
This alternative construction requires only one value of
 $\langle \tau \rangle$, but leads to the same form of 
the mass matrices, hence the same phenomenology.

The so constructed models involve three real parameters 
fixed by the values of the three charged lepton masses.
The three neutrino masses, three neutrino mixing angles and 
three CPV phases are functions of altogether 2 real constants 
and two phases. In these models the neutrino mixing matrix 
is of trimaximal mixing form. In Case I 
it is given by the tri-bimaximal mixing matrix 
multiplied on the right by a unitary 
rotation in the 1-3 plane, 
which depends on one angle and one phase. 
In addition to successfully describing the charged lepton masses, 
neutrino mass-squared differences and the atmospheric and reactor neutrino
mixing angles $\theta_{23}$ and $\theta_{13}$,
these models predict the values of the lightest neutrino mass 
(i.e., the absolute neutrino mass scale), of the Dirac and 
Majorana CP violation (CPV) phases and correspondingly 
of the effective neutrinoless double beta decay Majorana mass, 
as well as  the existence of specific correlations between
i) the values of the solar neutrino mixing 
angle $\theta_{12}$ and the angle $\theta_{13}$,
ii) the values of the Dirac CPV phase $\delta$ 
and of the angle  $\theta_{23}$,
iii) the sum of the  neutrino masses and $\theta_{23}$,
iv) the neutrinoless double beta decay 
effective Majorana mass and $\theta_{23}$, and
v) between the two Majorana phases 
(Fig.~\ref{fig:corr}).
These predictions will be tested  
with future more precise neutrino oscillation data, 
with results from direct neutrino mass and 
neutrinoless double beta decay experiments, 
as well as with improved cosmological measurements.

\section*{Acknowledgements}

P.P.N. and S.T.P. would like to thank João Penedo and Arsenii Titov
for numerous fruitful discussions of finite modular group symmetries and their
application to the lepton flavour problem.
This work was supported in part  by 
by  MEXT KAKENHI grant No  15K05045 (M.T.),
by the  World Premier 
International Research Center Initiative (WPI Initiative, MEXT), 
Japan (S.T.P.) and by the European Union's Horizon 2020 research 
and innovation programme under the 
 Marie Sklodowska-Curie grant agreements No 674896 
(ITN Elusives) and No 690575 
(RISE InvisiblesPlus) and by the INFN program on Theoretical 
Astroparticle Physics (P.P.N. and S.T.P.).


\appendix

\section*{Appendix}

\section{Multiplication rule of $A_4$ group}
\label{sec:multiplication-rule}
We take 
\begin{align}
\begin{aligned}
S=\frac{1}{3}
\begin{pmatrix}
-1 & 2 & 2 \\
2 &-1 & 2 \\
2 & 2 &-1
\end{pmatrix},
\end{aligned}
\qquad 
\begin{aligned}
T=
\begin{pmatrix}
1 & 0& 0 \\
0 &\omega& 0 \\
0 & 0 & \omega^2
\end{pmatrix}, 
\end{aligned}
\end{align}
where $\omega=e^{i\frac{2}{3}\pi}$ for a triplet.
In this base,
the multiplication rule of the $A_4$ triplet is
\begin{align}
\begin{pmatrix}
a_1\\
a_2\\
a_3
\end{pmatrix}_{\bf 3}
\otimes 
\begin{pmatrix}
b_1\\
b_2\\
b_3
\end{pmatrix}_{\bf 3}
&=\left (a_1b_1+a_2b_3+a_3b_2\right )_{\bf 1} 
\oplus \left (a_3b_3+a_1b_2+a_2b_1\right )_{{\bf 1}'} \nonumber \\
& \oplus \left (a_2b_2+a_1b_3+a_3b_1\right )_{{\bf 1}''} \nonumber \\
&\oplus \frac13
\begin{pmatrix}
2a_1b_1-a_2b_3-a_3b_2 \\
2a_3b_3-a_1b_2-a_2b_1 \\
2a_2b_2-a_1b_3-a_3b_1
\end{pmatrix}_{{\bf 3}}
\oplus \frac12
\begin{pmatrix}
a_2b_3-a_3b_2 \\
a_1b_2-a_2b_1 \\
a_3b_1-a_1b_3
\end{pmatrix}_{{\bf 3}\  } \ , \nonumber \\
\nonumber \\
{\bf 1} \otimes {\bf 1} = {\bf 1} \ , \qquad &
{\bf 1'} \otimes {\bf 1'} = {\bf 1''} \ , \qquad
{\bf 1''} \otimes {\bf 1''} = {\bf 1'} \ , \qquad
{\bf 1'} \otimes {\bf 1''} = {\bf 1} \  .
\end{align}
More details are shown in the review~\cite{Ishimori:2010au,Ishimori:2012zz}.





\newpage
\begin{thebibliography}{99}
	
	\bibitem{Tanabashi:2018oca}
    K. Nakamura and S.T. Petcov in
	M.~Tanabashi {\it et al.} [Particle Data Group],
	Phys.\ Rev.\ D {\bf 98} (2018)   030001.
%
\bibitem{Capozzi:2018ubv}
 F.~Capozzi, E.~Lisi, A.~Marrone and A.~Palazzo,
  Prog.\ Part.\ Nucl.\ Phys.\  {\bf 102} (2018) 48
  [arXiv:1804.09678 [hep-ph]].
%

\bibitem{Altarelli:2010gt}
G.~Altarelli and F.~Feruglio,
Rev.\ Mod.\ Phys.\  {\bf 82} (2010) 2701
[arXiv:1002.0211 [hep-ph]].



\bibitem{Ishimori:2010au}
H.~Ishimori, T.~Kobayashi, H.~Ohki, Y.~Shimizu, H.~Okada and M.~Tanimoto,
Prog.\ Theor.\ Phys.\ Suppl.\  {\bf 183} (2010) 1
[arXiv:1003.3552 [hep-th]].



\bibitem{Ishimori:2012zz}
H.~Ishimori, T.~Kobayashi, H.~Ohki, H.~Okada, Y.~Shimizu and M.~Tanimoto,
Lect.\ Notes Phys.\  {\bf 858} (2012) 1, Springer.


\bibitem{King:2013eh}
S.~F.~King and C.~Luhn,
Rept.\ Prog.\ Phys.\  {\bf 76} (2013) 056201
[arXiv:1301.1340 [hep-ph]].

\bibitem{King:2014nza}
S.~F.~King, A.~Merle, S.~Morisi, Y.~Shimizu and M.~Tanimoto,
arXiv:1402.4271 [hep-ph].

\bibitem{Tanimoto:2015nfa}
M.~Tanimoto,
AIP Conf.\ Proc.\  {\bf 1666} (2015) 120002.

\bibitem{Petcov:2017ggy}
S.~T.~Petcov,
Eur.\ Phys.\ J.\ C {\bf 78} (2018) no.9,  709
[arXiv:1711.10806 [hep-ph]].


\bibitem{Ma:2001dn}
E.~Ma and G.~Rajasekaran,
Phys.\ Rev.\  D {\bf 64}, 113012 (2001)
[arXiv:hep-ph/0106291].



\bibitem{Babu:2002dz}
K.~S.~Babu, E.~Ma and J.~W.~F.~Valle,
Phys.\ Lett.\  B {\bf 552}, 207 (2003)
[arXiv:hep-ph/0206292].

\bibitem{Altarelli:2005yp}
G.~Altarelli and F.~Feruglio,
Nucl.\ Phys.\ B {\bf 720} (2005) 64
[hep-ph/0504165].


\bibitem{Altarelli:2005yx}
G.~Altarelli and F.~Feruglio,
Nucl.\ Phys.\ B {\bf 741} (2006) 215
[hep-ph/0512103].

\bibitem{Shimizu:2011xg}
Y.~Shimizu, M.~Tanimoto and A.~Watanabe,
Prog.\ Theor.\ Phys.\  {\bf 126} (2011) 81
[arXiv:1105.2929 [hep-ph]].

\bibitem{Kang:2018txu}
S.~K.~Kang, Y.~Shimizu, K.~Takagi, S.~Takahashi and M.~Tanimoto,
PTEP {\bf 2018} (2018) no.8,  083B01
doi:10.1093/ptep/pty080
[arXiv:1804.10468 [hep-ph]].

\bibitem{Feruglio:2017spp}
F.~Feruglio,
arXiv:1706.08749 [hep-ph].

	
\bibitem{deAdelhartToorop:2011re} 
R.~de Adelhart Toorop, F.~Feruglio and C.~Hagedorn,
Nucl.\ Phys.\ B {\bf 858}, 437 (2012)
[arXiv:1112.1340 [hep-ph]].

\bibitem{Kobayashi:2018vbk}
T.~Kobayashi, K.~Tanaka and T.~H.~Tatsuishi,
Phys.\ Rev.\ D {\bf 98} (2018) no.1,  016004
[arXiv:1803.10391 [hep-ph]].


\bibitem{Penedo:2018nmg}
J.~T.~Penedo and S.~T.~Petcov,
arXiv:1806.11040 [hep-ph].

\bibitem{Criado:2018thu}
J.~C.~Criado and F.~Feruglio,
arXiv:1807.01125 [hep-ph].

\bibitem{Kobayashi:2018scp}
	T.~Kobayashi, N.~Omoto, Y.~Shimizu, K.~Takagi, 
M.~Tanimoto and T.~H.~Tatsuishi,
	arXiv:1808.03012 [hep-ph].

\bibitem{Novichkov:2018ovf}
P.~P.~Novichkov, J.~T.~Penedo, S.~T.~Petcov and A.~V.~Titov,
arXiv:1811.04933 [hep-ph].


\bibitem{Kobayashi:2018bff}
T. Kobayashi and S. Tamba, 
arXiv:1811.11384 [hep-th].

\bibitem{Novichkov:2018xyzw}
P.~P.~Novichkov, J.~T.~Penedo, S.~T.~Petcov and A.~V.~Titov,
arXiv:1812.02158 [hep-ph].



\bibitem{Ding:2013hpa}
G.~J.~Ding, S.~F.~King, C.~Luhn and A.~J.~Stuart,
JHEP {\bf 1305} (2013) 084
doi:10.1007/JHEP05(2013)084
[arXiv:1303.6180 [hep-ph]].


\bibitem{Ding:2013bpa}
G.~J.~Ding, S.~F.~King and A.~J.~Stuart,
JHEP {\bf 1312} (2013) 006
[arXiv:1307.4212 [hep-ph]].

\bibitem{Li:2014eia}
C.~C.~Li and G.~J.~Ding,
JHEP {\bf 1508} (2015) 017
doi:10.1007/JHEP08(2015)017
[arXiv:1408.0785 [hep-ph]].

\bibitem{Petcov:2014laa}
  S.~T.~Petcov,
  Nucl.\ Phys.\ B {\bf 892} (2015) 400
  [arXiv:1405.6006 [hep-ph]].

\bibitem{Girardi:2015rwa}
I.~Girardi, S.~T.~Petcov, A.~J.~Stuart and A.~V.~Titov,
Nucl.\ Phys.\ B {\bf 902} (2016) 1
[arXiv:1509.02502 [hep-ph]].

\bibitem{Girardi:2016zwz}
I.~Girardi, S.~T.~Petcov and A.~V.~Titov,
Nucl.\ Phys.\ B {\bf 911} (2016) 754
[arXiv:1605.04172 [hep-ph]].

\bibitem{Petcov:2018snn}
S.~T.~Petcov and A.~V.~Titov,
Phys.\ Rev.\ D {\bf 97} (2018) no.11,  115045
[arXiv:1804.00182 [hep-ph]].



		
\bibitem{BPont57} B. Pontecorvo, 
                  Zh. Eksp. Teor. Fiz. 
{\bf 33} (1957) 549 and {\bf 34} (1958) 247.
%

	\bibitem{Maki:1962mu}
	Z.~Maki, M.~Nakagawa and S.~Sakata,
	Prog.\ Theor.\ Phys.\  {\bf 28} (1962) 870.
	
	\bibitem{Pontecorvo:1967fh}
	B.~Pontecorvo,
	Sov.\ Phys.\ JETP {\bf 26} (1968) 984
	[Zh.\ Eksp.\ Teor.\ Fiz.\  {\bf 53} (1967) 1717].

	 
	\bibitem{Grimus:2008tt}
	W.~Grimus and L.~Lavoura,
	JHEP {\bf 0809} (2008) 106
	[arXiv:0809.0226 [hep-ph]].
	
	\bibitem{Albright:2010ap}
	C.~H.~Albright, A.~Dueck and W.~Rodejohann,
	Eur.\ Phys.\ J.\ C {\bf 70} (2010) 1099
	[arXiv:1004.2798 [hep-ph]].
	

	
	\bibitem{Shimizu:2014ria}
	Y.~Shimizu, M.~Tanimoto and K.~Yamamoto,
	Mod.\ Phys.\ Lett.\ A {\bf 30} (2015) 1550002
	[arXiv:1405.1521 [hep-ph]].
%
\bibitem{deAnda:2018ecu}
  F.~J.~de Anda, S.~F.~King and E.~Perdomo,
  arXiv:1812.05620 [hep-ph].
%
\bibitem{Baur:1901xyz} A.~Baur {\it et al.}, arXiv:1901.03251. 
%
\bibitem{Kobayashi:2018wkl} T. Kobayashi {\it et al.}, 
 arXiv:1812.11072 [hep-ph].

\end{thebibliography}
\end{document}